\documentclass[prd,aps,showpacs,nofootinbib,tightenlines]{revtex4}
\usepackage{mathrsfs}
\usepackage{amsmath}
\usepackage{amssymb}
\usepackage{epsfig}
\usepackage{graphicx}
\usepackage{booktabs}
\usepackage{multirow}
\usepackage{subfigure}
\usepackage{bm}
\usepackage{times}
\usepackage{braket}
\usepackage{color}
\usepackage{slashed}
\usepackage{hyperref}
\definecolor{Red}{rgb}{1.,0.,0.}

\definecolor{Blue}{rgb}{0.,0.,1.}

\definecolor{nicered}{rgb}{0.7,0.1,0.1}
\definecolor{nicegreen}{rgb}{0.1,0.5,0.1}
\hypersetup{colorlinks,citecolor=nicegreen,linkcolor=nicered}
\def \cpc{ Chin. Phys. C  }

\def \epjc{ Eur. Phys. J. C }

\def \jpg{  J. Phys. G }

\def \plb{  Phys. Lett. B }
\def \ppnp{ Prog. Part. Nucl. Phys. }
\def \prd{  Phys. Rev. D }
\def \prl{  Phys. Rev. Lett.  }

\def \jhep{ J. High Energy Phys. }
\begin{document}

\newcommand{\beq}{\begin{eqnarray}}
\newcommand{\eeq}{\end{eqnarray}}
\newcommand{\non}{\nonumber\\ }

\title{Quasi-two-body decays $B_{(s)}\to P f_2(1270)\to P\pi\pi$ in the perturbative QCD approach}

\author{Ya Li$^1$}                \email [Corresponding author: ] {liyakelly@163.com}
\author{Ai-Jun Ma$^2$}           \email []{theoma@163.com}
\author{Zhou Rui$^3$}              \email [] {jindui1127@126.com}
\author{Wen-Fei Wang$^4$}         \email [] {wfwang@sxu.edu.cn}
\author{Zhen-Jun Xiao$^{5}$}    \email []  {xiaozhenjun@njnu.edu.cn}
\affiliation{$^1$ Department of Physics, College of Science, Nanjing Agricultural University, Nanjing, Jiangsu 210095, P.R. China}
\affiliation{$^2$ Department of Mathematics and Physics, Nanjing Institute of Technology, Nanjing, Jiangsu 211167, P.R. China}
\affiliation{$^3$ College of Sciences, North China University of Science and Technology,
                          Tangshan 063009,  P.R. China}
\affiliation{$^4$ Institute of Theoretical Physics, Shanxi University, Taiyuan, Shanxi 030006, P.R. China}
\affiliation{$^5$ Department of Physics and Institute of Theoretical Physics,
                          Nanjing Normal University, Nanjing, Jiangsu 210023, P.R. China}

\date{\today}

\begin{abstract}
  In this work, we calculate  the $CP$-averaged branching ratios and direct $CP$-violating asymmetries of the quasi-two-body decays $B_{(s)}\to P f_2(1270)\to P\pi\pi$ with the two-pion  distribution amplitude $\Phi_{\pi\pi}^{\rm D}$ by using the perturbative QCD factorization approach, where $P$ represents a light pseudoscalar meson $K, \pi, \eta$ and $\eta^{\prime}$.
  The relativistic Breit-Wigner formula for the $D$-wave resonance $f_2(1270)$ is adopted to parameterize the timelike form factor $F_{\pi}$, which contains the final state interactions between the pions in the resonant regions.
  The consistency of theoretical results with data can be achieved by determining the Gegenbauer moments of the $D$-wave two-pion distribution amplitudes.
  The decay rates for the considered decay modes are generally in the order of $10^{-9}$ to $ 10^{-6}$.
  The integrated direct $CP$ asymmetries for the charged modes agree with the {\it BABAR} and Belle measurements.
  As a by-product, we extract the branching ratios of $B_{(s)}\to Pf_2(1270)$ from the corresponding quasi-two-body decay modes, which still need experimental tests at the ongoing and forthcoming experiments.
\end{abstract}

\pacs{13.25.Hw, 12.38.Bx}

\maketitle

\newpage
\section{Introduction}
Three-body hadronic $B$ meson decays are a rich field for studying the direct $CP$ violation and testing the standard model and QCD.
Recent measurements by {\it BABAR}~\cite{prd83-112010,prd79-072006,prd78-052005,prd78-012004,prd80-112001,prd76-031101} and Belle~\cite{prd75-012006,prl96-251803,prd71-092003} Collaborations of a number of $B \to \pi\pi\pi$, $B \to K\pi\pi$, or $B \to J/\psi \pi\pi$ decays have triggered considerable theoretical interests in understanding three-body hadronic $B$ decays.
These three-body decays are known experimentally to be dominated by the low energy resonances
on $\pi\pi$, $KK$ and $K\pi$ channels.
As the LHCb Collaboration reported, the $\pi\pi$ final states are found to comprise the decay products of the $\rho(770)$, $f_0(500)$, $f_0(980)$, $f_2(1270)$\footnote{ For the sake of simplicity, we
generally use the abbreviation $f_2=f_2(1270)$ in the following sections. }, and $f_0(1370)$ (etc.) mesons in case of $B^0$ or $B^0_s$ decays~\cite{npb871-403,prd87-052001,prd90-012003,prd86-052006,prd89-092006,prd95-012006}.

It is known that such three-body $B$ decay modes are more intractable than those two-body decays due to the entangled resonant and nonresonant contributions, as well as the possible final-state interactions (FSIs)~\cite{prd89-094013,1512-09284,89-053015}, whereas the relative strength of these contributions vary significantly for different decay modes.
The analysis of these three-body decays utilizing the Dalitz plots \cite{dalitz-plot1,dalitz-plot2} enables one to investigate the properties of various tensor, vector, and scalar resonances with the isobar model~\cite{123-333,prd11-3165} in terms of the usual Breit-Wigner (BW) model~\cite{BW-model}.
Unfortunately, no proof of factorization has been given for the decays of the $B$ into three mesons.
However, we can restrict ourselves to specific kinematical configurations, in which two
energetic final state mesons almost collimate to each other.
Then the three-body interactions are expected to be suppressed strongly in such conditions.
It seems reasonable to assume the validity of factorization for these quasi-two-body $B$ decays.
In the ``quasi-two-body" mechanism, the two-body scattering and all possible interactions
between the two involved  particles are included but the interactions between the third
particle and the pair of mesons are neglected.

There are several theoretical approaches for describing hadronic three-body decays of $B$ mesons based on the symmetry principles and factorization theorems.
$U$-spin and flavor $SU(3)$ symmetries were adopted in Refs.~\cite{prd72-094031,plb727-136,prd72-075013,prd84-056002,plb728-579,prd91-014029}.
The QCD-improved factorization (QCDF)~\cite{npb675-333} has been widely applied to the studies of three-body hadronic $B$ meson decays in Refs.~\cite{npb899-247,plb622-207,prd74-114009,prd79-094005,APPB42-2013,prd76-094006,prd88-114014,prd94-094015,prd89-094007,prd87-076007}.
The authors investigated the detailed factorization properties of the $B^+\to\pi^+\pi^+\pi^-$ mode in different regions of phase space~\cite{npb899-247}.
In Ref.~\cite{APPB42-2013}, the authors studied the decays of $B^\pm \to \pi^\pm \pi^\mp \pi^\pm$ within a quasi-two-body
QCD factorization approach and introduced the scalar and vector form factors for the $S$ and $P$ waves, as well as a relativistic BW formula for the $D$ wave to describe the meson-meson final state interactions.
In Ref.~\cite{prd94-094015}, for instance, the authors studied the nonresonant contributions using heavy meson chiral perturbation theory~\cite{prd46-1148,prd45-2188,plb280-287} with some modifications and analyzed
the resonant contributions with the isobar model by using the usual BW formalism.

Relying on the perturbative QCD (PQCD) approach, furthermore, some three-body $B$ meson decays have also been investigated in  Refs.~\cite{Chen:2002th,Chen:2004th,prd97-034033,1803-02656}.
The authors of Ref.~\cite{prd89-074031} studied the direct $CP$ asymmetries of $B^\pm \to
\pi^\pm \pi^+ \pi^-$ and $ K^\pm \pi^+\pi^-$ decays by fitting the time-like form factors and the rescattering phases contained in the two-pion
distribution amplitudes $\Phi_{h_1h_2}$~\cite{MP,MT01,MT02,MT03,MN,Grozin01,Grozin02} to relevant experimental data.
Very recently, in the PQCD approach, we studied the $S$-wave resonance contributions to the
decays $B^0_{(s)}\to J/\psi\pi^+\pi^-$~\cite{prd91-094024}, $B^0_{(s)}\to \eta_c{(1S,2S)}\pi^+\pi^-$ \cite{epjc76-675,cpc41-083105},
$B^0_{s}\to \psi(2S,3S) \pi^+\pi^-$~\cite{epjc77-199} and $B^0_{(s)}\to J/\psi(\psi(2S)) K \pi$~\cite{prd97-033006}, as well as the $P$-wave resonance contributions to the decays
$B \to P (\rho,\rho(1450),\rho(1700)) \to P \pi\pi$~\cite{plb763-29,prd95-056008,prd96-036014}, $B \to D (\rho,\rho(1450),\rho(1700)) \to D \pi\pi$~\cite{npb923-54,1708-01889,1710-00327} and $B \to \eta_c(1S,2S) (\rho,\rho(1450),\rho(1700)) \to \eta_c(1S,2S)\pi\pi$~\cite{npb924-745}.
One of the aims for studying such three-body $B$ meson decays is to test the usability of our PQCD approach.
The above works do support our general expectation: the PQCD factorization is universal for exclusive hadronic three-body $B$ meson decays.

In the PQCD factorization approach, the contribution from the dynamical region, where there is at least one pair of the final state light mesons
having an invariant mass below $O(\bar\Lambda m_B)$~\cite{Chen:2002th}, $\bar\Lambda=m_B-m_b$ being the $B$ meson and $b$ quark mass difference, is dominant.
Because the hard $b$-quark decay kernels containing two hard gluons at leading order is not important due to the power-suppression, the configuration involving two energetic mesons almost parallel to each other may provide the dominant contribution.
Then it's reasonable that the dynamics associated with the pair of mesons can be factorized into a two-meson distribution
amplitude $\Phi_{h_1h_2}$.
The typical PQCD factorization formula for the $B\to h_1h_2h_3$ decay amplitude can be described as the form of ~\cite{Chen:2002th}
\begin{eqnarray}
\mathcal{A}=\Phi_B\otimes H\otimes \Phi_{h_1h_2}\otimes\Phi_{h_3},
\end{eqnarray}
where the hard kernel $H$ contains only one hard gluon and describes the dynamics of the strong and electroweak interactions in the three-body hadronic decays as in the formalism for the two-body $B$ meson decays.
The $\Phi_B$ and $\Phi_{h_3}$ are the wave functions for the $B$ meson
and the final state $h_3$, which absorb the non-perturbative dynamics in the relevant  processes.
In the PQCD approach based on the $k_T$ factorization theorem, we adopt the widely used wave function for $B$ meson~\cite{prd63-054008}, which includes the intrinsic $b$ dependence with $b$ being a variable conjugate to $k_T$.
In Ref.~\cite{JHEP06-013}, the authors pointed out that the operator-level definition of the transverse-momentum-dependent (TMD)
hadronic wave functions is highly nontrivial in order to avoid the potential light-cone divergence and the rapidity singularity.
A well-defined TMD can be found in Ref.~\cite{JHEP02-008}.
Meanwhile, the Sudakov factors from the $k_T$ resummation have been included to suppress the long-distance contributions from the large $b$
region in this work.
The more precise joint resummation derived in~\cite{JHEP01-004} can be included in the future.
For the QCD resummation, one can include its effect as going beyond the tree level in PQCD analysis, which will be done in the future by taking into account the results as given in Refs.~\cite{JHEP02-008,JHEP01-004}.

\begin{figure}[tbp]
\vspace{-1cm}
\centerline{\epsfxsize=10cm \epsffile{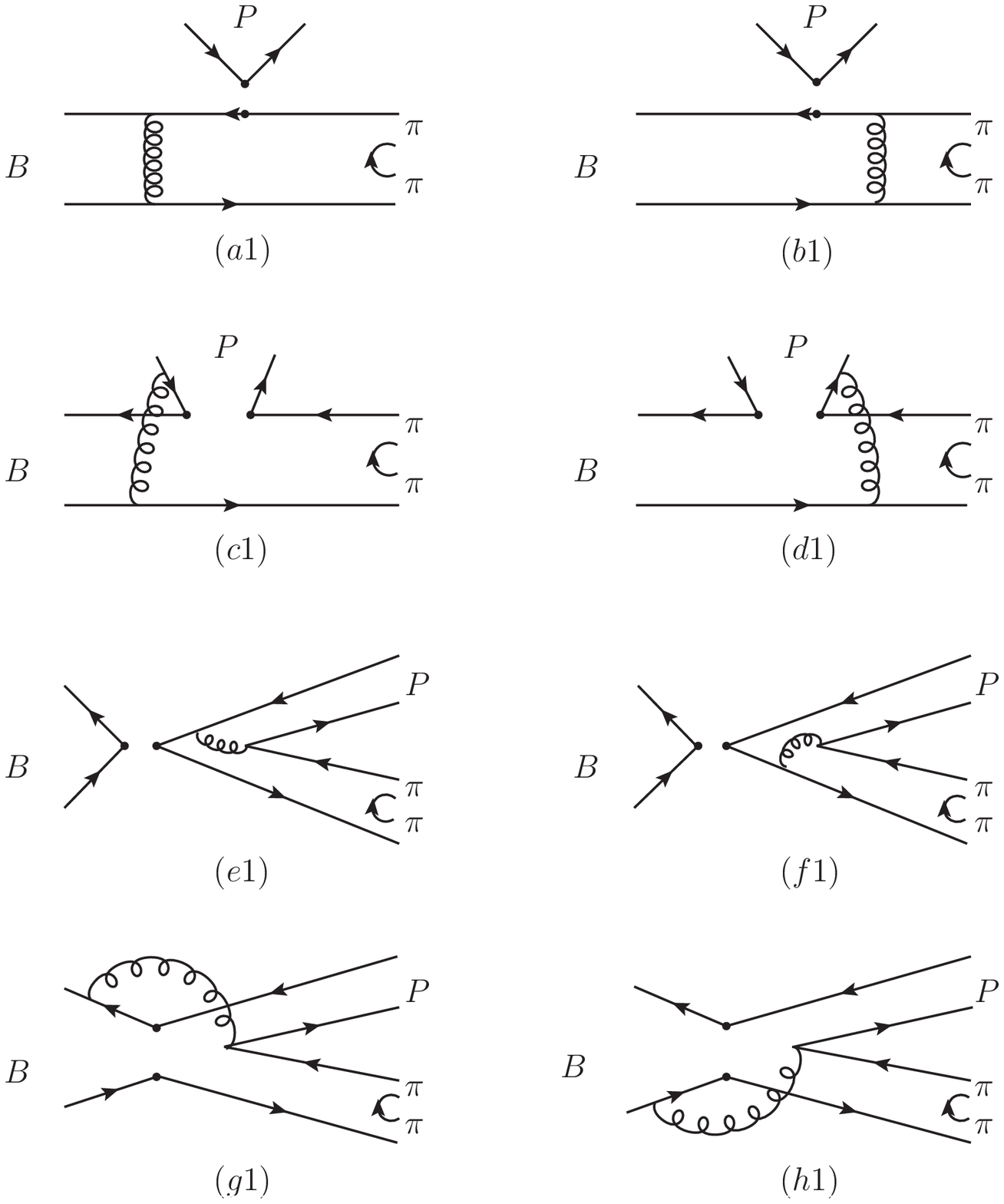}}
\caption{Typical Feynman diagrams for the quasi-two-body decays $B \to P(f_2(1270) \to) \pi \pi$,
where $B$ stands for the $B^\pm, B^0$ or $B_s$ meson and $P$ denotes $K, \pi, \eta$ or $\eta^\prime$.
With $\alpha=a$-$d$ and $\beta=e$-$h$, the diagrams ($\alpha$1) for the
$B\to f_2(1270)\to\pi\pi$ transition and the diagrams ($\beta$1) for annihilation contributions.}
\label{fig-fig1}
\end{figure}

\begin{figure}[tbp]
\centerline{\epsfxsize=10cm \epsffile{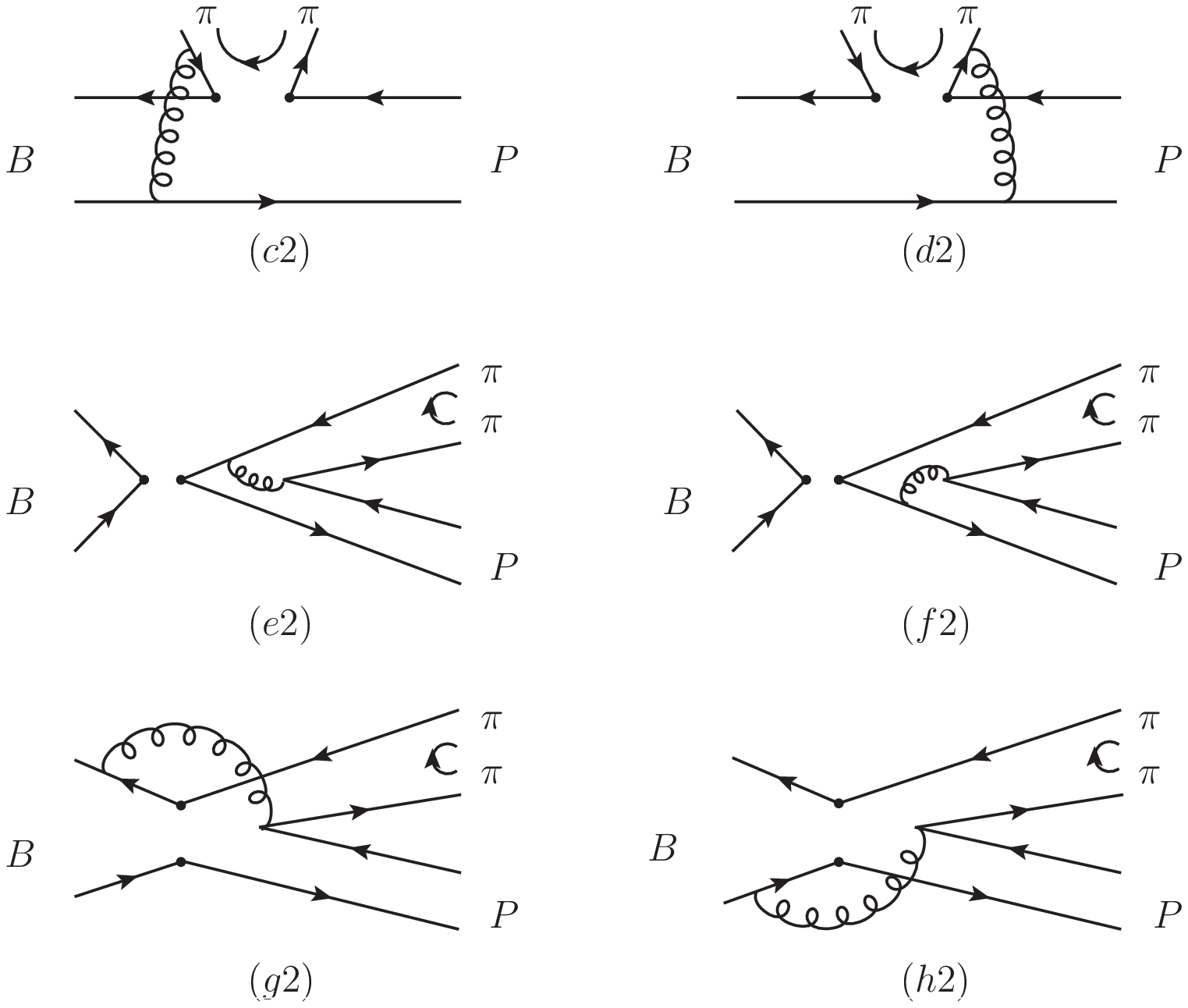}}
\caption{Typical Feynman diagrams for the quasi-two-body decays $B \to P(f_2(1270) \to )\pi \pi$,
where $B$ stands for the $B^\pm, B^0$ or $B_s$ meson and $P$ denotes $K, \pi, \eta$ or $\eta^\prime$.
With $\alpha=c$-$d$ and $\beta=e$-$h$, the diagrams ($\alpha$2) for the $B\to P$ transition and
the diagrams ($\beta$2) for annihilation contributions.}
\label{fig-fig2}
\end{figure}

In this work, we will extend our previous work on $S$ and $P$- wave resonances to the $D$-wave ones in the PQCD framework. Taking the decays $B_{(s)}\to P f_2(1270)\to P\pi\pi$, $P=(\pi, K, \eta$ or $\eta^\prime)$ as examples,
the relevant Feynman diagrams are illustrated in Fig.~\ref{fig-fig1} and ~\ref{fig-fig2}.
Since  the tensor resonance  cannot be created from the $(V\pm A)$, $(S\pm P)$ or tensor current,
the diagrams with  a  $D$-wave $\pi\pi$ pair emitted in Fig.~\ref{fig-fig2} are prohibited in naive factorization.
Phenomenologically there are growing appeals for the
two-body charmless hadronic $B$ decays involving a light tensor meson in the  final states in the past few years \cite{prd67-014002,prd83-014007,prd82-054019,prd83-034001,prd83-014008,prd86-094015,plb732-36,prd96-013005}.
More recently, one of us has investigated the two-body  decays of the $B_c$ meson into
the tensor charmonium using the PQCD approach~\cite{prd97-033001}.
Experimentally the $CP$-averaged branching ratios and direct $CP$-violating asymmetries of quasi-two-body decays $B \to \pi f_2(1270)\to \pi\pi\pi$~\cite{prd79-072006} and $B \to K f_2(1270) \to K \pi\pi$~\cite{prd78-012004,prd80-112001,prl96-251803,prd71-092003} have been measured.
One can see that the measured $CP$ violation is just a number in two-body $B$ decays, while the $CP$ asymmetry depends on the invariant mass displayed in the Dalitz plot in the three-body modes~\cite{prd90-112004}.
It is meaningful to study the decays  $B \to P f_2(1270) \to P \pi\pi$ in the three-body framework, which provide useful information for understanding the $CP$-violation mechanisms. 
For the $D$-wave resonant state $f_2(1270)$,  the relativistic BW formula is adopted to parametrize the timelike form factors $F_{\pi}$, which contains the final state interactions between the pions in the resonant regions.
The agreement of theoretical results with data can be achieved by determining the appropriate Gegenbauer moments of the $D$-wave two-pion distribution amplitudes.
Just like the $\eta$-$\eta^{\prime}$ mixing in the pseudoscalar case, the isoscalar tensor states $f_2(1270)$ and $f_2^{\prime}(1525)$ also have a similar mixing.
The mixing angle between the $f_2(1270)$ and $f_2^{\prime}(1525)$ is really small due to a fact that the $f_2(1270)$($f_2^{\prime}(1525)$) predominantly decays into $\pi\pi$ ($K\bar{K}$).
In our paper, we focus on the resonances on $\pi\pi$ channel and leave the detailed  discussion about mixtures of $f_2(1270)$-$f_2^{\prime}(1525)$ for future studies associated with precise experimental measurements.

The present paper is organized as follows.
In Sec.~II, we give a brief introduction for the theoretical framework.
The numerical values, some discussions and the conclusions will be given in last two sections.
The explicit PQCD factorization formulas for all the decay amplitudes are collected in the Appendix.

\section{FRAMEWORK}\label{sec:2}
In the light-cone coordinates, the $B$ meson momentum $p_{B}$, the total momentum of the pion pair,
$p=p_1+p_2$, the momentum $p_3$ of the final state meson $P$, and the momentum $k_B$ of the spectator quark in the $B$ meson,
the momentum $k$ for the resonant state $f_2$, $k_3$ for the final state $P$ are in the form of
\begin{eqnarray}\label{mom-pBpp3}
p_{B}&=&\frac{m_{B}}{\sqrt2}(1,1,0_{\rm T}),~\quad p=\frac{m_{B}}{\sqrt2}(1,\eta,0_{\rm T}),~\quad
p_3=\frac{m_{B}}{\sqrt2}(0,1-\eta,0_{\rm T}),\nonumber\\
k_{B}&=&\left(0,x_B \frac{m_{B}}{\sqrt2} ,k_{B \rm T}\right),\quad k= \left( z\frac{m_{B}}{\sqrt2},0,k_{\rm T}\right),\quad
k_3=\left(0,(1-\eta)x_3 \frac{m_B}{\sqrt{2}},k_{3{\rm T}}\right), \label{mom-B-k}
\end{eqnarray}
where $m_{B}$ is the mass of $B$ meson, the variable $\eta$ is defined as $\eta=\omega^2/m^2_{B}$ with
the invariant mass squared $\omega^2=p^2=(p_1+p_2)^2$.
The parameters $x_B, z, x_3$ are chosen as the momentum fraction of the positive quark in each meson and run from zero to unity.
$k_{B \rm T}, k_{\rm T}$, and $k_{3{\rm T}}$ denote the transverse momentum of the positive quark, respectively.
We define $\zeta=p^+_1/p^+$ as one of the pion pair's momentum fraction, the two pions momenta
$p_{1,2}$ can be described as
\begin{eqnarray}
 p_1=\left ( \zeta \frac{m_B}{\sqrt{2}}, (1-\zeta)\eta \frac{m_B}{\sqrt{2}}, p_{1\rm T} \right),~\quad
 p_2=\left ( (1-\zeta)\frac{m_B}{\sqrt{2}}, \zeta\eta \frac{m_B}{\sqrt{2}}, p_{2\rm T} \right ),
\end{eqnarray}
where $p_{1\rm T}^2=p_{2\rm T}^2=\zeta(1-\zeta)\omega^2$.

We here adopt the $D$-wave two-pion distribution amplitude similar as the one being used in Ref.~\cite{plb763-29},
\begin{eqnarray}
\Phi_{\pi\pi}^{\rm D}=\frac{1}{\sqrt{2N_c}}\left [ { p \hspace{-2.0truemm}/ }\Phi_{v\nu=-}^{I=0}(z,\zeta,\omega^2)+\omega\Phi_{s}^{I=0}(z,\zeta,\omega^2)
+\frac{{p\hspace{-1.5truemm}/}_1{p\hspace{-1.5truemm}/}_2
  -{p\hspace{-1.5truemm}/}_2{p\hspace{-1.5truemm}/}_1}{w(2\zeta-1)}
  \Phi_{t\nu=+}^{I=0}(z,\zeta,\omega^2) \right ] \;.
\label{eq:phifunc}
\end{eqnarray}
For $I=0$, the distribution amplitude $\Phi_{v\nu=-}^{I=0}$ contributes at twist-2, $\Phi_s^{I=0}$ and $\Phi_{t\nu=+}^{I=0}$ contribute at twist-3.
It is worthwhile to stress that this $\pi$-$\pi$ system has similar asymptotic distribution amplitudes as the ones for a tensor meson~\cite{prd83-034001,prd83-014008,prd86-094015}, but replacing the tensor decay constants with the timelike form factor:
\begin{eqnarray}
\Phi_{v\nu=-}^{I=0}&=&\phi_0=\frac{6F_{\pi}(s)}{2\sqrt{2N_c}}z(1-z)\left[3 a^0_1(2z-1)\right] P_2(2\zeta-1) \;, \label{eq:phi1}\\
\Phi_{s}^{I=0}&=&\phi_s=-\frac{9F_s(s)}{4\sqrt{2N_c}}\left[a^0_1(1-6z+6z^2)\right] P_2(2\zeta-1) \;, \label{eq:phi2}\\
\Phi_{t\nu=+}^{I=0}&=&\phi_t=\frac{9F_t(s)}{4\sqrt{2N_c}}\left[a^0_1(1-6z+6z^2)(2z-1)\right] P_2(2\zeta-1) \;,
\label{eq:phi3}
\end{eqnarray}
where the Legendre polynomial $P_2(2\zeta-1)=1-6\zeta(1-\zeta)$.
The twist-3 distribution amplitudes should be fixed by the equations of motion~\cite{npb529-323,npb543-201} related to twist-2 ones.
The moment $a^0_1$ is regarded as a free parameter and determined in this work.

Now, we focus on the dipion electromagnetic form factor.
Taking the resonance contribution to the pion form factor into account, the pion electromagnetic form factor is defined in the standard way $\langle \pi^+(p_1)\pi^-(p_2)|j^{em}_{\mu}|0\rangle=(p_1-p_2)_{\mu}F_{\pi}(s)$, where $s=(p_1+p_2)^2$ is the timelike momentum transfer squared and $s\geq 4m^2_{\pi}$~\cite{epjc39-41}.
The form factor $F_{\pi}(s)$ can be analytically continued to the spacelike region $s<0$, corresponding to the hadronic matrix element $\langle \pi^+(p_1)|j^{em}_{\mu}|\pi^+(-p_2)\rangle$ by crossing-symmetry.
Even so, the continuation from the timelike to spacelike region does not work well for the resonance $J/\psi$ as shown in Ref.~\cite{epjc39-41}.
As is well known, the electromagnetic form factor of pion at large (spacelike)
momentum transfer on the basis of one-pion distribution amplitude has been computed in Ref.~\cite{prd83-054029} with the PQCD approach at NLO.
Applying the analytical continuation of the obtained
result in the kinetic variable of momentum-transfer squared, one
should be able to compute the timelike pion electromagnetic form
factor directly without resorting to fitting the experimental measurements~\cite{plb749-1}.
However, it's difficult to make the analytical continuation from the spacelike to timelike region for the dipion form factor.
When we start from the spacelike region, it is not easy to identify the decay width piece, which can be interpreted as contribution of multihadron states to the imaginary part of the form factor in the resonant contribution~\cite{epjc39-41,prd89-014015}.
In other word, it is impossible to generate the pole from the spacelike region by the analytical continuation.

Certainly, an alternative way to account for the hadronic resonance effect is that the electromagnetic form factor of pion at large (timelike) momentum transfer can be computed from perturbative QCD factorization approach at large momentum transfer with the parton-hadron duality ansatz.
In order to account for the hadronic resonance effect, one can apply the hadronic dispersion relation for the electromagnetic form factor of the pion in the entire kinematic region and then implement the constraints from QCD calculation at large $Q^2$ for the determination of the unknown hadronic parameters
entering the nonperturbative parametrization of the dispersion form.
We will make efforts to calculate the electromagnetic form factor of pion at large (timelike) momentum transfer in the future.
In this work, the relativistic BW formula for the $D$-wave resonance $f_2(1270)$ is adopted to parametrize the timelike form factor $F_{\pi}(s)$ and the
explicit simplified expressions are in the following form,
\begin{eqnarray}
F_{\pi}(s)&=&\frac{m_{f_2}^2}{m^2_{f_2} -s-im_{f_2}\Gamma(s)},\\
\Gamma(s)&=&\Gamma_{f_2}\left(\frac{\sqrt{s-4m^2_{\pi}}}{\sqrt{m^2_{f_2}-4m^2_{\pi}}}\right)^5\frac{m_{f_2}}{\sqrt{s}}
\frac{X_{J=2}(\frac{1}{2}\sqrt{s-4m^2_{\pi}})}{X_{J=2}(\frac{1}{2}\sqrt{m^2_{f_2}-4m^2_{\pi}})},\\
X_{J=2}(z)&=&\frac{1}{(zr_{BW})^4+3(zr_{BW})^2+9},
\end{eqnarray}
with the two-pion invariant mass squared $s=\omega^2=m^2(\pi\pi)$ and the $m_{f_2}=1.276 {\rm GeV}$ and $\Gamma_{f_2}=0.187{\rm GeV}$ are the pole mass and width of resonance state $f_2(1270)$, respectively.
We find that the variations of radius parameter $r_{BW}=4$~\cite{prd79-072006} do not significantly change the values in our calculations.
Hence, it is reasonable to set $r_{BW}=0$ in our latter numerical calculations.
The approximate relations $F_{s,t}(s)\approx (f_{f_2}^T/f_{f_2}) F_\pi(s)$~\cite{plb763-29} will also be used
in the following section.

\section{Numerical results}\label{sec:3}
\begin{table}
\caption{The decay constants of $f_2(1270)$ meson is from~\cite{prd82-054019}, while other parameters are adopted in PDG~\cite{pdg2016}
in  our numerical calculations.  }
\label{tab:constant1}
\begin{tabular*}{14.5cm}{@{\extracolsep{\fill}}lccccccc}
  \hline\hline
Masses (\text{GeV}) &&& $m_{B}=5.280$ & $m_{B_s}=5.367$ & $m_{b}=4.66$ & $m_{f_2(1270)}=1.276$ &$m_{\pi^{\pm}}=0.140$ \\[1ex]
&& &$m_{\pi^0}=0.135$ &$m_{K^{\pm}}=0.494$ &$m_{K^0}=0.498$ &$m_{\eta}=0.548$ &$m_{\eta^{\prime}}=0.958$     \\[1ex]
\end{tabular*}
\begin{tabular*}{14.5cm}{@{\extracolsep{\fill}}lcccc}
{{The
Wolfenstein parameters}} &$A=0.811 \quad \lambda=0.22506$ \quad $\bar{\rho} = 0.124$ \quad $\bar{\eta}= 0.356$\\[1ex]
\end{tabular*}
\begin{tabular*}{14.5cm}{@{\extracolsep{\fill}}lcccc}
Decay constants (MeV) & $f_{B}=190.9\pm 4.1$& $f_{B_s}=227.2\pm 3.4$ & $f_{f_2(1270)}=102\pm{6}$ &$f^T_{f_2(1270)}=117\pm{25}$\\[1ex]
\end{tabular*}
\begin{tabular*}{14.5cm}{@{\extracolsep{\fill}}lccc}
Lifetimes (ps) & $\tau_{B_s}=1.51$& $\tau_{B_0}=1.52$& $\tau_{B^+}=1.638$\\[1ex]
\hline\hline
\end{tabular*}
\end{table}

For the numerical calculations, those parameters such as meson masses, the Wolfenstein parameters, decay constants, and the lifetimes
of $B_{(s)}$ mesons are given in Table \ref{tab:constant1},
while the $B$ meson and kaon (pion) distribution amplitudes are the same as widely adopted in the PQCD approach~\cite{prd95-056008,epjc28-515,li2003,prd85-094003}.

For the decays $B \to P f_2(1270) \to P \pi \pi$, the differential branching ratio is written as~\cite{pdg2016},
\begin{eqnarray}
\frac{d{\cal B}}{ds}=\tau_{B}\frac{|\overrightarrow{p_{\pi}}||\overrightarrow{p_3}|}{64\pi^3m^3_{B}}|{\cal A}|^2. \label{expr-br}
\end{eqnarray}
The kinematic variables $|\overrightarrow{p_{\pi}}|$ and $|\overrightarrow{p_3}|$ denote one of the
pion pair's and $P$'s momentum in the center-of-mass frame of the pion pair,
\begin{eqnarray}
 |\overrightarrow{p_{\pi}}|=\frac12\sqrt{s-4m^2_{\pi}}, \quad~~
 |\overrightarrow{p_3}|=\frac12  \sqrt{\big[(m^2_{B}-M_3^2)^2-2(m^2_{B}+M_3^2) s+s^2 \big]/s}. \label{br-momentum}
\end{eqnarray}

\begin{table}[!!b]
\caption{$CP$ averaged branching ratios of $B_{(s)}\to P (f_2(1270) \to) \pi^+ \pi^-$ decays calculated in PQCD approach together
with experimental data~\cite{pdg2016}. The theoretical errors correspond to the uncertainties due to the shape parameters $\omega_{B_{(s)}}$ in the wave function of $B_{(s)}$ meson, the Gegenbauer moment $a_1^0$ and the next-to-leading-order effects (the hard scale $t$ and the QCD scale $\Lambda_{\rm QCD}$), respectively.}
 \label{Presults}
\begin{center}
\begin{tabular}{c c c c}
 \hline \hline
 \multicolumn{1}{c}{}&\multicolumn{2}{c}{ Quasi-two-body $\mathcal {B}$ (in $10^{-7}$)} &\multicolumn{1}{c}{}  \\
{Modes}           & Scenario I~  &Scenario II   ~   &Experiment \footnotemark[1]   \\
\hline\hline
$B^+ \to K^+(f_2(1270)\to)\pi^+ \pi^-$             &$11.09^{+1.60+6.23+2.82}_{-1.45-4.85-3.18}$  &$12.77^{+1.80+7.18+3.22}_{-1.62-5.59-3.61}$ &$6.01\pm{1.52}$\\

 $B^0 \to K^0(f_2(1270)\to)\pi^+ \pi^-$            &$8.81^{+1.35+4.95+2.48}_{-1.09-3.86-2.56}$   &$10.30^{+1.54+5.79+2.85}_{-1.21-4.51-2.90}$ &$~~15.16^{+7.30}_{-6.74}~~$\\

 $B^0_s \to \bar{K}^0(f_2(1270)\to)\pi^+ \pi^-$    &$0.37^{+0.04+0.20+0.06}_{-0.04-0.16-0.11}$   &$0.42^{+0.05+0.24+0.10}_{-0.04-0.18-0.12}$   &$-$\\

 \hline
  $B^+ \to \pi^+(f_2(1270)\to)\pi^+ \pi^-$         &$10.55^{+2.06+5.93+0.90}_{-1.70-4.62-0.89}$   &$10.49^{+2.05+5.89+0.87}_{-1.70-4.59-0.88}$    &$~~8.98^{+3.93}_{-2.25}~~$\\

  $B^0 \to \pi^0(f_2(1270)\to)\pi^+ \pi^-$          &$0.30^{+0.03+0.17+0.03}_{-0.04-0.13-0.06}$  &$0.33^{+0.04+0.18+0.03}_{-0.05-0.15-0.06}$    &$-$ \\   

  $B^0_s \to \pi^0(f_2(1270)\to)\pi^+ \pi^-$        &$0.003^{+0.000+0.002+0.001}_{-0.001-0.001-0.001}$  &$0.008^{+0.002+0.002+0.001}_{-0.001-0.003-0.001}$   &$-$\\

 \hline
  $B^0 \to \eta(f_2(1270)\to)\pi^+ \pi^-$           &$0.52^{+0.08+0.29+0.07}_{-0.07-0.23-0.09}$  &$0.52^{+0.08+0.29+0.06}_{-0.07-0.23-0.09}$    &$-$\\

  $B_s^0 \to \eta(f_2(1270) \to)\pi^+ \pi^-$        &$1.35^{+0.43+0.75+0.23}_{-0.33-0.59-0.46}$  &$1.78^{+0.63+1.01+0.32}_{-0.45-0.78-0.50}$     &$-$\\

  $B^0 \to \eta^{\prime}(f_2(1270)\to)\pi^+ \pi^-$  &$0.61^{+0.12+0.35+0.08}_{-0.09-0.27-0.11}$  &$0.63^{+0.12+0.35+0.08}_{-0.10-0.28-0.13}$    &$-$\\

  $B_s^0 \to \eta^{\prime}(f_2(1270)\to)\pi^+ \pi^-$  &$2.70^{+0.73+1.52+0.48}_{-0.58-1.18-0.88}$ &$4.83^{+1.33+2.72+0.88}_{-1.03-2.11-1.33}$    &$-$\\

 \hline \hline
\end{tabular}
\end{center}
\footnotetext[1]{ The experimental results are obtained by multiplying the relevant measured two-body branching ratios according to the Eq.~(\ref{eq:def1}). }
\end{table}

\begin{table}[t!!]
\caption{Direct $CP$-violating asymmetries of $B_{(s)}\to P (f_2(1270) \to) \pi^+ \pi^-$ decays calculated in PQCD approach together
with experimental data~\cite{pdg2016}. The theoretical errors correspond to the uncertainties due to the shape parameters $\omega_{B_{(s)}}$ in the wave function of $B_{(s)}$ meson and the next-to-leading-order effects (the hard scale $t$ and the QCD scale $\Lambda_{\rm QCD}$), respectively.}
 \label{PPresults}
\begin{center}
\begin{tabular}{p{5cm}p{3.5cm}p{3cm} c }
 \hline \hline
 \multicolumn{1}{c}{}&\multicolumn{2}{c}{ ${\cal A}_{\rm CP} (\%) $}&\multicolumn{1}{c}{} \\
{Modes}           & Scenario I~  &Scenario II  &  Experiment     \\
\hline\hline
$B^+ \to K^+(f_2(1270)\to)\pi^+ \pi^-$      &$-48.2^{+1.3+12.4}_{-0.9-13.9}$    &$-45.6^{+1.4+11.6}_{-0.8-12.3}$    &$-68^{+19}_{-17}$      \\
 $B^0 \to K^0(f_2(1270)\to)\pi^+ \pi^-$     &$1.1^{+0.7+0.9}_{-0.6-0.0}$            &$1.1^{+0.5+0.8}_{-0.4-0.1}$     &$-$\\
 $B^0_s \to \bar{K}^0(f_2(1270)\to)\pi^+ \pi^-$    &$-39.3^{+4.9+1.4}_{-5.1-2.9}$    &$-37.0^{+4.5+0.8}_{-5.2-3.0}$       &$-$ \\
 \hline
  $B^+ \to \pi^+(f_2(1270)\to)\pi^+ \pi^-$   &$28.6^{+1.2+5.9}_{-3.1-4.4}$           &$28.9^{+1.1+6.0}_{-3.2-4.7}$      &$41\pm{30}$  \\
  $B^0 \to \pi^0(f_2(1270)\to)\pi^+ \pi^-$   &$-19.7^{+2.1+17.8}_{-5.1-11.7}$          &$-17.2^{+1.6+16.4}_{-4.5-12.3}$     &$-$  \\
  $B^0_s \to \pi^0(f_2(1270)\to)\pi^+ \pi^-$  &$-0.2^{+0.0+25.9}_{-8.0-37.3}$ &$-13.5^{+6.1+24.6}_{-0.0-0.7}$ &$-$\\
 \hline
  $B^0 \to \eta(f_2(1270)\to)\pi^+ \pi^-$    &$-65.1^{+2.5+10.2}_{-0.1-20.2}$     &$-64.4^{+1.5+10.5}_{-0.6-19.7}$         &$-$   \\
  $B_s^0 \to \eta(f_2(1270) \to)\pi^+ \pi^-$  &$-1.6^{+0.2+2.1}_{-0.3-1.5}$    &$2.0^{+0.0+1.4}_{-0.9-0.8}$  &$-$\\
  $B^0 \to \eta^{\prime}(f_2(1270)\to)\pi^+ \pi^-$    &$-28.3^{+2.6+0.2}_{-2.8-2.1}$      &$-28.0^{+3.1+0.6}_{-2.3-1.2}$          &$-$ \\
  $B_s^0 \to \eta^{\prime}(f_2(1270)\to)\pi^+ \pi^-$  &$2.1^{+0.3+1.9}_{-0.0-0.0}$   &$4.6^{+1.0+0.7}_{-0.6-0.2}$ &$-$\\
 \hline \hline
\end{tabular}
\end{center}
\end{table}
By using the differential branching fraction in Eq.~(\ref{expr-br}), and the decay amplitudes in the Appendix, we calculate the $CP$ averaged branching ratios ($\cal B$) and direct $CP$-violating asymmetries ($\cal A_{CP}$) for the decays $B_{(s)}\to P (f_2 \to) \pi \pi $.
In this work, we consider two scenarios for the tensor meson $f_2(1270)$.
In scenario I, it is assumed that $f_2(1270)$ is a pure nonstrange isospin
singlet state $(u\bar{u}+d\bar{d})/\sqrt{2}$,
while in scenario II, the strange state $s\bar{s}$ enters the contributions
with a nonvanishing mixing angle just like the $\eta-\eta^{\prime}$ mixing in the pseudoscalar sector.
Thus, the isoscalar tensor states can be written as
\begin{eqnarray}
f_2(1270)=\frac{1}{\sqrt{2}}\left(u\bar{u}+d\bar{d}\right)\cos\theta_{f_2}+\left(s\bar{s}\right)\sin\theta_{f_2}\;,\non
f_2^{\prime}(1525)=\frac{1}{\sqrt{2}}\left(u\bar{u}+d\bar{d}\right)\sin\theta_{f_2}-\left(s\bar{s}\right)\cos\theta_{f_2}\;.
\label{eq:mix}
\end{eqnarray}
The detailed discussions about the mixing angle could be found in Refs.~\cite{prd84-094008,jpg27-807}.
Here we employ the most recent updated value $(9\pm1)^\circ$  from PDG2016~\cite{pdg2016}.
The predictions on the $CP$ averaged branching ratios and direct $CP$-violating asymmetries from scenario I and II are enumerated distinctly
in Tables~\ref{Presults} and \ref{PPresults}, as well as the current available data, respectively.
The fit to the data~\cite{pdg2016} determines the Gegenbauer moment $a^0_1=0.40$, which differs
from that in the distribution amplitudes for a longitudinally polarized $f_2(1270)$ meson~\cite{prd83-034001,prd83-014008,prd86-094015}.

In our numerical calculations, the first theoretical uncertainty results from the variations of the shape parameter $\omega_{B_{(s)}}$ of the $B_{(s)}$ meson distribution amplitude.  We adopt the value $\omega_B=0.40\pm0.04$~GeV or $\omega_{B_s}=0.50\pm0.05$~GeV and vary its value with a 10\% range, which is supported by intensive PQCD studies~\cite{prd63-054008,epjc23-275,prd63-074009,plb504-6}.
It is shown that the shape parameter $\omega_B$ can reach about 20\% in magnitude for the main uncertainties.
We note that another value $\omega_B (1 {\rm GeV})=0.354^{+0.038}_{-0.030}$~GeV implied by the light-cone sum rules calculations of the semileptonic $B \to \pi$ form factors with $B$-meson DAs on the light-cone~\cite{npb898-563} has been taken in the Refs.~\cite{JHEP09-159,JHEP05-184}.
This number is very close to our error range, which result in the branching ratios changing 20 percents as mentioned above.
Model-independent determinations of the inverse moment of the $B$-meson light-cone distribution amplitude $\omega_B$ in HQET have been discussed extensively from
the radiative leptonic $B$-meson decays with the QCD factorization approach and the dispersion relations in Refs.~\cite{JHEP09-159,JHEP05-184}.
The opportunity of constraining the inverse moment $\omega_B$ should be explored with the improvement
of better measurements at the Belle II experiment in the near future.
The second error comes from the Gegenbauer moment $a^0_1=0.40\pm0.10$.
The last one is caused by the variation of the hard scale $t$ from $0.75t$ to $1.25t$ (without changing $1/b_i$) and the QCD scale $\Lambda_{\rm QCD}=0.25\pm0.05$~GeV, which characterizes the effect of the NLO QCD contributions.
For the $CP$ averaged branching ratios, the second error contributes the main uncertainties in our approach,
while the other two errors are comparable and less than $30\%$.
For the direct $CP$-violating asymmetries, the error from the Gegenbauer moment
is  largely cancelled between the  numerator and denominator, and the main uncertainty refer to the hard scale.
The uncertainties from $\tau_{B^\pm}$, $\tau_{B^0}$, $\tau_{B_s}$ and the Wolfenstein parameters in~\cite{pdg2016} are small and have been neglected.
The significance of the radius parameter $r_{BW}$ has been verified in our calculations.
Taking the decay channel $B^+ \to K^+(f_2\to)\pi^+\pi^-$ in scenario I as an example, the branching ratio $\mathcal{B}=11.09 \times 10^{-7}$ and
direct $CP$ asymmetry ${\cal A_{CP}}=-48.2\%$ for $r_{BW}=0$, while for $r_{BW}=4.0$, the corresponding values are $\mathcal{B}=10.77\times 10^{-7}$ and ${\cal A_{CP}}=-48.4\%$.
One can find that the difference between the results with two choices is really small.

It is observed that the branching ratios from the two scenarios are comparable for the same $B$ meson decay modes, while in the case of $B_s$ decays, the two scenarios are rather  different.
For example, the value of $\mathcal{B}(B_s \to \eta^{\prime}(f_2 \to) \pi^+\pi^-)$ in scenario II are nearly twice as that in scenario I.
The main reason is that the factorizable contributions from the $P$ emission diagrams [see Fig.~\ref{fig-fig1}(a1) and Fig.~\ref{fig-fig1}(b1)] in $s\bar{s}$ component
enhance the corresponding decay amplitudes.
However, for the $B$ meson decays, such factorizable contributions come from the $n\bar{n}$ ($n=u,d$) component, while the $s\bar{s}$ will contribute to the annihilation diagrams or nonfactorizable emission diagrams, which are power suppressed when
compared with the factorizable emission diagrams according to the power counting rules in the factorization assumption.

It is shown that $f_2(1270)$ is primarily an $(u\bar{u}+d\bar{d})/\sqrt{2}$ state for the case of $B$ meson decay modes, whereas for the $B_s$ decays, the $s\bar{s}$ component makes a significant contributions albeit with suffering a large suppression from the mixing angle, especially for the process $B_s \to \eta^{\prime}(f_2 \to) \pi^+\pi^-$.
Therefore, we recommend the LHCb and/or Belle-II experiments to measure this mode to probe the precise structures of $f_2(1270)$.
In addition, from Table~\ref{Presults}, it is found that ${\cal B }(B^0_{(s)} \to \eta(f_2\to)\pi^+ \pi^-)<{\cal B }(B^0_{(s)} \to \eta^{\prime}(f_2\to)\pi^+ \pi^-)$.
Since both $\eta_q$ and $\eta_s$ will contribute in these modes, but the relative sign of the $\eta_q$ state with respect
to the $\eta_s$ state is negative for $\eta$ and positive for $\eta^{\prime}$, which leads to destructive interference between $\eta_q$ and $\eta_s$ for the former, but constructive interference for the latter.

Combined with the Clebsch-Gordan Coefficients, we can describe the relation
\begin{eqnarray}
|\pi\pi \rangle =\frac{1}{\sqrt{3}}|\pi^+\pi^-\rangle+\frac{1}{\sqrt{3}}|\pi^-\pi^+\rangle-\frac{1}{\sqrt{3}}|\pi^0\pi^0\rangle.
\end{eqnarray}
Isospin conservation is assumed for the strong decays of an $I=0$ resonance $f_2$ to $\pi\pi$ when we compute the branching fraction of the quasi-two-body process $B\to Pf_2$, namely,
\begin{eqnarray}
\frac{\Gamma(f_2 \to \pi^+\pi^-)}{\Gamma(f_2 \to \pi\pi)}=2/3.
\end{eqnarray}
According to the relation of the decay rates between the quasi-two-body and the corresponding two-body decay modes
\begin{eqnarray}
\mathcal{B}( B_{(s)} \to P (f_2 \to ) \pi^+ \pi^- ) =
\mathcal{B}( B_{(s)} \to P f_2 ) \cdot {\mathcal B}(f_2 \to\pi\pi)\cdot \frac{2}{3}, \label{eq:def1}
\end{eqnarray}
with ${\mathcal B}(f_2 \to\pi\pi)=(84.2^{+2.9}_{-0.9})\%$~\cite{pdg2016}, we further obtain the PQCD predictions for $ {\cal B}( B/B_s \to P f_2)$ from the values as listed in the second column of Table~\ref{Presults}:
\begin{eqnarray}
\mathcal {B}(B^+ \to K^+ f_2)&=&[19.76^{+12.51}_{-10.65}]\times 10^{-7}\;, \non
\mathcal {B}(B^0 \to K^0 f_2)&=&[15.69^{+10.15}_{-8.48}]\times 10^{-7}\;, \non
\mathcal {B}(B^0_s \to \bar{K}^0 f_2)&=&[0.66^{+0.37}_{-0.36}]\times 10^{-7}\;,\non
\mathcal {B}(B^+ \to \pi^+ f_2)&=&[18.79^{+11.29}_{-8.91}]\times 10^{-7}\;, \non
\mathcal {B}(B^0 \to \pi^0 f_2)&=&[0.53^{+0.32}_{-0.27}]\times 10^{-7}\;,\non
\mathcal {B}(B^0_s \to \pi^0 f_2)&=&[0.005\pm0.004]\times 10^{-7}\;, \non
\mathcal {B}(B^0 \to \eta f_2)&=&[0.93^{+0.55}_{-0.46}]\times 10^{-7}\;, \non
\mathcal {B}(B^0_s \to \eta f_2)&=&[2.40^{+1.59}_{-1.46}]\times 10^{-7}\;, \non
\mathcal {B}(B^0 \to \eta^{\prime} f_2)&=&[1.09^{+0.68}_{-0.55}]\times 10^{-7}\;, \non
\mathcal {B}(B^0_s \to \eta^{\prime} f_2)&=&[4.81^{+3.12}_{-2.81}]\times 10^{-7}\;,
\end{eqnarray}
where the individual errors have been added in quadrature.
\begin{figure}[tbp]
\centerline{\epsfxsize=8.5cm \epsffile{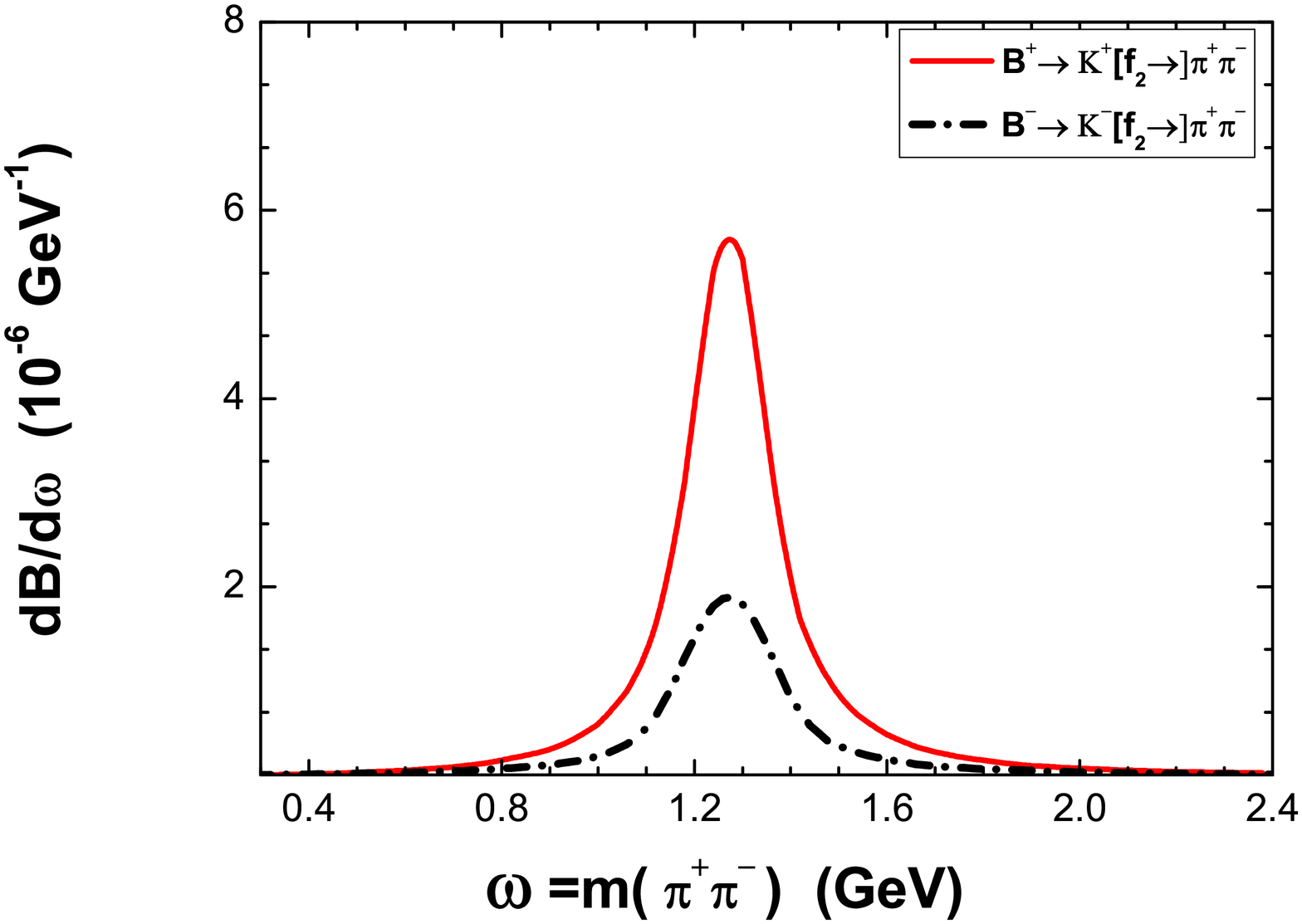}
            \epsfxsize=8.5cm \epsffile{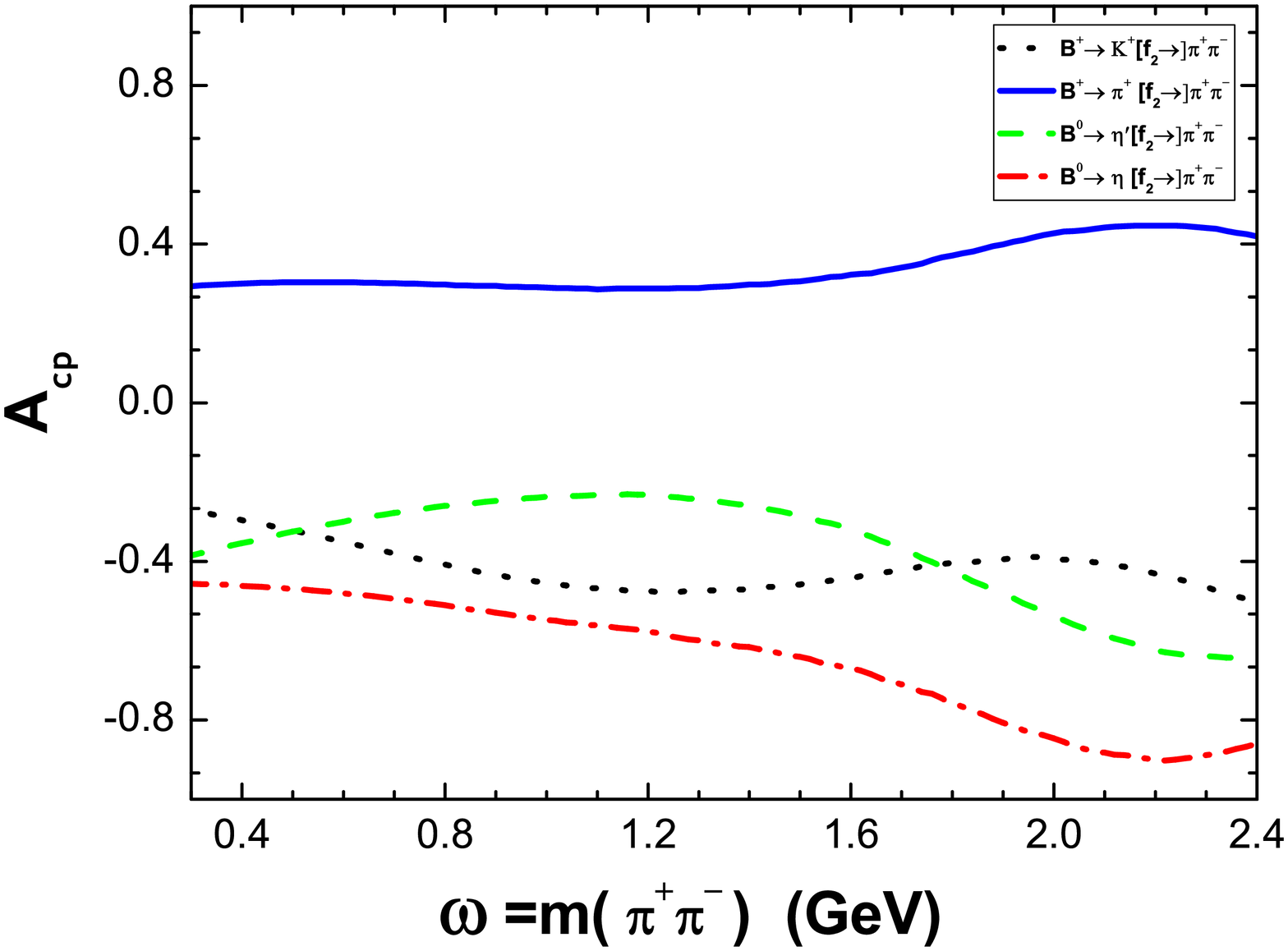}}
\vspace{-0.2cm}
  {\scriptsize\bf (a)\hspace{8cm}(b)}
\caption{(a) Differential branching ratios for the $B^\pm\to K^\pm f_2(1270)\to K^\pm\pi^+\pi^-$ decays.
(b) Differential distribution of ${\cal A}_{cp}$ in $\omega$ for the decay modes $B^+ \to K^+[f_2\to]\pi^+\pi^-$, $B^+ \to \pi^+[f_2\to]\pi^+\pi^-$, $B^0 \to \eta^{\prime}[f_2\to]\pi^+\pi^-$ and $B^0 \to \eta[f_2\to]\pi^+\pi^-$.}
\label{fig-br}
\end{figure}

One can see that some channels have both large branching ratios and direct $CP$ asymmetries, letting the corresponding
measurement appear feasible.
In fact, some physical observables (such as the $CP$ averaged branching
ratios and  direct  $CP$ violations) of the two charged  decay modes like $B^+ \to K^+(f_2 \to)\pi^+\pi^-$ and $B^+ \to \pi^+(f_2 \to) \pi^+\pi^-$
have been searched by ${\it BABAR}$  and Belle Collaborations  \cite{prd79-072006,prd78-012004,prl96-251803,prd71-092003}.
For example,  ${\it BABAR}$ Collaboration  \cite{prd79-072006} reported  a measurement,
$\mathcal{B}( B^+ \to \pi^+(f_2 \to) \pi^+ \pi^-)=(0.90^{+0.37}_{-0.24})\times10^{-6}$,
which agrees with our calculations in both scenarios I and II.
Furthermore, for the $K$ mode, the measurements from  ${\it BABAR}$ \cite{prd78-012004} and Belle \cite{prl96-251803} are the following
\begin{eqnarray}
\mathcal {B}(B^+ \to K^+(f_2 \to)\pi^+ \pi^-)=\left\{
\begin{aligned}
(&0.50^{+0.21}_{-0.19})\times 10^{-6}, \quad\quad  &\text{{\it BABAR}}, \nonumber\\
(&0.75 ^{+0.21}_{-0.25})\times 10^{-6}, \quad\quad  &\text{Belle}.  \nonumber\\
\end{aligned}\right.
\end{eqnarray}
Their weighted average,  enter the numbers given in Table~\ref{Presults},
 are typically smaller than our prediction. None the less, taking
the errors into consideration, the theoretical prediction and
experimental data can still agree with each other.
For the direct $CP$ asymmetries, although the error bars from the data are still large,
we are happy to see that all these measured entries have the same sign as our theoretical calculations [see Table~\ref{PPresults}].

From Table~\ref{Presults}, one can  see that the branching ratio of $B^+ \to K^+(f_2 \to)\pi^+ \pi^-$ decay is a little larger than that of $B^0 \to K^0(f_2 \to)\pi^+ \pi^-$ decay due to the extra contribution from the tree diagrams [see Eq.~(\ref{amp1}) and ~(\ref{amp2})] and the larger lifetime of the $B^+$ meson for the former decay mode.
The similar situations also appear  in the previous calculations from  QCDF~\cite{prd83-034001} and PQCD~\cite{prd86-094015}.
However, the data~\cite{pdg2016} shows that the latter decay mode has a relative large decay rate.
It is worth of noting that this mode also has much larger relative errors because of limited statistics and the Dalitz-plot signal model dependence~\cite{prd80-112001}. Such a difference should be clarified in the forthcoming experiments based on much larger data samples.

According to the full Dalitz-plot analysis to the $B^\pm \to \pi^\pm \pi^\pm\pi^\mp$ decay by the
{\it BABAR} experiment~\cite{prd79-072006}, the dominant contributions come from the $P$-wave
resonance $\rho(770)$ and nonresonant contributions.
The relative rate between the contributions from the $P$-wave resonance $\rho(770)$ and $D$-wave resonance $f_2(1270)$ was measured to be
\begin{eqnarray}
R_{exp} \equiv \frac{{\cal B}(B^\pm \to \pi^\pm (\rho^0(770) \to) \pi^+\pi^-)}{{\cal B}(B^\pm
\to \pi^\pm (f_2(1270) \to) \pi^+\pi^-)} = 9.00^{+0.59}_{-1.48}. 
\end{eqnarray}
For a more direct comparison with this available experimental data,
we use our previous PQCD calculation  of the $P$-wave resonance contribution  $\mathcal {B}(B^+ \to \pi^+(\rho^0(770) \to)\pi^+ \pi^-)=(8.84^{+1.91}_{-1.69})\times 10^{-6}$, where all errors are combined in quadrature, as an input.
Combined with the prediction on $\mathcal {B}(B^+\to \pi^+ (f_2(1270) \to) \pi^+\pi^-)$ in Table~\ref{Presults},
we obtain the ratio
 \begin{eqnarray}
R_{PQCD}= \frac{{\cal B}(B^+ \to \pi^+ (\rho^0(770) \to) \pi^+\pi^-)}{{\cal B}(B^+
\to \pi^+ (f_2(1270) \to) \pi^+\pi^-)} = 8.38^{+4.50}_{-2.02},
\end{eqnarray}
which is consistent with above {\it BABAR} data quite well.
For the similar ratio for the $K$ counterpart,  the calculated value is $3.64^{+2.91}_{-0.93}$,
which is compatible with {\it BABAR} data $R_{exp}=7.12^{+2.30}_{-1.08}$~\cite{prd78-012004} within about two standard deviations and Belle data $R_{exp}=5.19^{+1.35}_{-0.47}$~\cite{prl96-251803} within about one standard deviation.
These results suggest that the PQCD factorization approach is suitable for describing
the quasi-two-body $B$ meson decays through analyzing various resonances by reconstructing $\pi\pi$ final states
and  reproducing the invariant mass spectra of Dalitz plots.

Different from the fixed kinematics of the two-body decays, the decay amplitudes  of
the quasi-two-body decays depends on the $\pi\pi$ invariant mass,
which resulting in the differential distribution of branching ratios and direct $CP$ asymmetries.
In Fig.~\ref{fig-br}(a), we plot the differential branching ratios of the $B^\pm\to K^\pm f_2\to K^\pm\pi^+\pi^-$ decays.
One can see that the differential branching ratios of the $B^\pm\to K^\pm f_2\to K^\pm\pi^+\pi^-$ decays exhibit peaks at the $f_2$ meson mass.
Thus, the main portion of the branching ratios lies in the region around the pole mass of the $f_2(1270)$ resonance as expected.
The central values of the branching ratio ${\cal B}$ are $6.03\times 10^{-7}$ and $8.51\times 10^{-7}$ when the integration over $\omega$ is limited in the range of $\omega=[m_{f_2}-0.5\Gamma_{f_2}, m_{f_2}+0.5\Gamma_{f_2}]$ or
$\omega=[m_{f_2}-\Gamma_{f_2}, m_{f_2}+\Gamma_{f_2}]$ respectively, which amount to
$54\%$ and $77\%$ of the total branching ratio ${\cal B}=11.09\times10^{-7}$ as listed in Table~\ref{Presults}.
In Fig.~\ref{fig-br}(b), we display the differential distributions of ${\mathcal A}_{CP}$ for the four decay modes $B^+ \to K^+[f_2\to]\pi^+\pi^-$ (black dotted line), $B^+ \to \pi^+[f_2\to]\pi^+\pi^-$ (blue solid line), $B^0 \to \eta^{\prime}[f_2\to]\pi^+\pi^-$ (green dashed line), and $B^0 \to \eta[f_2\to]\pi^+\pi^-$(red dash-dotted line), respectively.
One can find a falloff of ${\mathcal A}_{CP}$ with $\omega$ for
$B^+ \to K^+[f_2\to]\pi^+\pi^-$, $B^0 \to \eta^{\prime}[f_2\to]\pi^+\pi^-$,
and $B^0 \to \eta[f_2\to]\pi^+\pi^-$.
It implies that the direct $CP$ asymmetries in the above three quasi-two-body
decays, if calculated as the two-body decays with the $f_2$ resonance mass being fixed to $m_{f_2}$, may be overestimated.
The ascent of the differential distribution of ${\mathcal A}_{CP}$ with $\omega$
for $B^+ \to \pi^+[f_2\to]\pi^+\pi^-$ suggests that its direct $CP$ asymmetry,
if calculated in the two-body formalism, may be underestimated.
In two-body $B$ decays, the measured $CP$ violation is just a number.
But in three-body decays, one can measure the distribution of $CP$ asymmetry in the Dalitz plot.
Hence, the Dalitz-plot analysis of ${\mathcal A}_{CP}$ distributions can reveal very rich information about $CP$ violation.
In the future, we will make more efforts to describe the distributions of $CP$ asymmetries for various resonances in the Dalitz plot.

\section{CONCLUSION}

In this work, we calculated the quasi-two-body decays $B_{(s)} \to  P f_2(1270) \to  P \pi\pi$ with $ P =(\pi,K,\eta,\eta^\prime )$ by utilizing the timelike form factor $F_\pi(s)$ within the PQCD factorization framework.
The relativistic Breit-Wigner formula for the $D$-wave resonance $f_2(1270)$ was adopted to parametrize the timelike form factors $F_{\pi}$, which contains the final state interactions between the pions in the resonant regions.
Using the determined Gegenbauer moments of the $D$-wave two-pion distribution amplitudes,
we have predicted the branching ratios and the direct $CP$ asymmetries of the
$B_{(s)}\to Pf_2\to P\pi\pi$ channels, and compared their differential branching ratios with currently available data.
General agreements between the PQCD predictions and the data could be achieved, although there is no enough data at present.
We have taken two scenarios of constituents of $f_2(1270)$ into consideration and found that the interference between $(u\bar{u}+d\bar{d})/\sqrt{2}$ and $s\bar{s}$ part can result in remarkable effects on some decay modes.
The branching ratios of the corresponding two-body decays have been extracted from the quasi-two-body decay modes.
More precise data from the LHCb and the future Belle II will test our predictions.

\begin{acknowledgments}

Many thanks to Hsiang-nan Li, Zhi-Tian Zou, and Qin Qin for valuable discussions.
This work was supported by the National Natural Science Foundation of
China under Grants No.~11775117, No.~11547038, No.~11605060, and No.~11547020.
\end{acknowledgments}

\appendix

\section{Decay amplitudes}

When the meson $f_2(1270)$ is treated as an pure $\frac{1}{\sqrt{2}}(u\bar{u}+d\bar{d})$ state, the total decay amplitude for each considered decay mode in this work are given as follows:
\begin{eqnarray}
{\cal A}(B^+ \to K^+(f_2 \to)\pi^+ \pi^-) &=& \frac{G_F} {2} \big\{V_{ub}^*V_{us}[(\frac{C_1}{3}+C_2)(F^{LL}_{ef_2}+F^{LL}_{af_2})+C_1(M^{LL}_{ef_2}+M^{LL}_{af_2})+C_2M^{LL}_{eP}]\nonumber\\
&-&V_{tb}^*V_{ts}[(\frac{C_3}{3}+C_4+\frac{C_9}{3}+C_{10})(F^{LL}_{ef_2}+F^{LL}_{af_2})+(\frac{C_5}{3}+C_6+\frac{C_7}{3}+C_8)(F^{SP}_{ef_2}+F^{SP}_{af_2})\nonumber\\
&+&(C_3+C_9)(M^{LL}_{ef_2}+M^{LL}_{af_2})+(C_5+C_7)(M^{LR}_{ef_2}+M^{LR}_{af_2})+(2C_4+\frac{C_{10} }{2}) M^{LL}_{eP}\nonumber\\
&+&(2C_6+\frac{ C_8}{2}) M^{SP}_{eP}]\big\} \;, \label{amp1}
\end{eqnarray}
\begin{eqnarray}
{\cal A}(B^0 \to K^0(f_2 \to)\pi^+ \pi^-) &=& \frac{G_F} {2}\big\{V_{ub}^*V_{us}[C_2 M^{LL}_{eP}]-V_{tb}^*V_{ts}[(\frac{C_3}{3}+C_4-\frac{1}{2}(\frac{C_9}{3}+C_{10}))(F^{LL}_{ef_2}+F^{LL}_{af_2})\nonumber\\
&+&(\frac{C_5}{3}+C_6-\frac{1}{2}(\frac{C_7}{3}+C_8))(F^{SP}_{ef_2}+F^{SP}_{af_2})+(C_3-\frac{C_9}{2})(M^{LL}_{ef_2}+M^{LL}_{af_2})\nonumber\\
&+&(C_5-\frac{C_7}{2})(M^{LR}_{ef_2}+M^{LR}_{af_2})+(2C_4+\frac{C_{10}}{2})M^{LL}_{eP}+(2C_6+\frac{C_8}{2})M^{SP}_{eP}]\big\} \;,\label{amp2}
 \end{eqnarray}
\begin{eqnarray}
{\cal A}(B_s^0 \to \bar{K}^0(f_2 \to)\pi^+ \pi^-) &=& \frac{G_F} {2}\big\{V_{ub}^*V_{ud}[C_2 M^{LL}_{eP}]-V_{tb}^*V_{td}[(C_3+2C_4-\frac{C_9}{2}+\frac{C_{10}}{2})M^{LL}_{eP}+(2C_6+\frac{ C_8}{2})M^{SP}_{eP}\nonumber\\
&+&(C_5-\frac{C_7}{2})(M^{LR}_{eP}+M^{LR}_{aP})+(\frac{C_3}{3}+C_4-\frac{1}{2}(\frac{C_9}{3}+C_{10}))F^{LL}_{aP}\nonumber\\
&+&(\frac{C_5}{3}+C_6-\frac{1}{2}(\frac{C_7}{3}+C_8))F^{SP}_{aP}+(C_3-\frac{C_9}{2})M^{LL}_{aP}]\big\} \;,\label{amp3}
\end{eqnarray}
\begin{eqnarray}
{\cal A}(B^+ \to \pi^+(f_2 \to)\pi^+ \pi^-) &=& \frac{G_F} {2}\big\{V_{ub}^*V_{ud}[(\frac{C_1}{3}+C_2)(F^{LL}_{ef_2}+F^{LL}_{af_2}+F^{LL}_{aP})
+C_1 (M^{LL}_{ef_2}+M^{LL}_{af_2}+M^{LL}_{aP})+C_2 M^{LL}_{eP}]\nonumber\\
&-&V_{tb}^*V_{td}[(\frac{C_3}{3}+C_4+\frac{C_9}{3}+C_{10})(F^{LL}_{ef_2}+F^{LL}_{af_2}+F^{LL}_{aP})+(\frac{C_5}{3}+C_6+\frac{C_7}{3}+C_8)(F^{SP}_{ef_2}\nonumber\\
&+&F^{SP}_{af_2}+F^{SP}_{aP})+(C_3+C_9)(M^{LL}_{ef_2}+M^{LL}_{af_2}+M^{LL}_{aP})+(C_5+C_7)(M^{LR}_{ef_2}+M^{LR}_{af_2}+M^{LR}_{aP})\nonumber\\
&+&(C_3+2C_4-\frac{C_9}{2}+\frac{C_{10}}{2})M^{LL}_{eP}+(C_5-\frac{C_7}{2})M^{LR}_{eP}+(2C_6+\frac{C_8}{2})M^{SP}_{eP}]\big\} \;,\label{amp4}
\end{eqnarray}
\begin{eqnarray}
 {\cal A}(B^0 \to \pi^0(f_2 \to)\pi^+ \pi^-) &=& \frac{G_F} {2\sqrt{2}}
 \big\{V_{ub}^*V_{ud}[(C_1+\frac{C_2}{3})(F^{LL}_{ef_2}+F^{LL}_{af_2}+F^{LL}_{aP})+C_2(M^{LL}_{ef_2}+M^{LL}_{af_2}-M^{LL}_{eP}+M^{LL}_{aP})]\nonumber\\
&-&V_{tb}^*V_{td}[(-\frac{C_3}{3}-C_4-\frac{3}{2}(C_7+\frac{C_8}{3})+\frac{5 C_9}{3}+C_{10})(F^{LL}_{ef_2}+F^{LL}_{af_2}+F^{LL}_{aP}) \nonumber\\
&+&(-C_3+\frac{C_9}{2}+\frac{3C_{10}}{2})(M^{LL}_{ef_2}+M^{LL}_{af_2}+M^{LL}_{aP})+\frac{3C_8}{2}(M^{SP}_{ef_2}+M^{SP}_{af_2}+M^{SP}_{aP})\nonumber\\
&-&(\frac{C_5}{3}+C_6-\frac{C_7}{6}-\frac{C_8}{2})(F^{SP}_{ef_2}+F^{SP}_{af_2}+F^{SP}_{aP})-(C_3+2C_4-\frac{C_9}{2}+\frac{C_{10}}{2})M^{LL}_{eP}\nonumber\\
&-&(C_5-\frac{C_7}{2})(M^{LR}_{ef_2}+M^{LR}_{af_2}+M^{LR}_{eP}+M^{LR}_{aP})-(2C_6+\frac{C_8}{2})M^{SP}_{eP}]\big\} \;,\label{amp5}
\end{eqnarray}
 \begin{eqnarray}
{\cal A}(B_s^0 \to \pi^0(f_2 \to)\pi^+ \pi^-) &=& \frac{G_F} {2\sqrt{2}}
\big\{V_{ub}^*V_{us}[(C_1+\frac{C_2}{3})(F^{LL}_{af_2}+F^{LL}_{aP})+C_2(M^{LL}_{af_2}+M^{LL}_{aP})]
-V_{tb}^*V_{ts}[\frac{3}{2}(C_9+\frac{C_{10}}{3}\nonumber\\
&-&C_7-\frac{C_8}{3})(F^{LL}_{af_2}+F^{LL}_{aP})+\frac{3C_{10}}{2}(M^{LL}_{af_2}+M^{LL}_{aP})+\frac{3C_8}{2}(M^{SP}_{af_2}+M^{SP}_{aP})]\big\} \;,\label{amp6}
\end{eqnarray}
\begin{eqnarray}
{\cal A}(B^0 \to \eta_q(f_2 \to)\pi^+ \pi^-) &=& \frac{G_F} {2\sqrt{2}}
 \big\{V_{ub}^*V_{ud}[(C_1+\frac{C_2}{3})(F^{LL}_{ef_2}+F^{LL}_{af_2}+F^{LL}_{aP})+C_2( M^{LL}_{ef_2}+M^{LL}_{af_2}
+M^{LL}_{eP}+M^{LL}_{aP})]\nonumber\\
&-&V_{tb}^*V_{td}[(\frac{7 C_3}{3}+\frac{5 C_4}{3}-2(C_5+\frac{C_6}{3})-\frac{1}{2}(C_7+\frac{C_8}{3}-\frac{2}{3}(C_9-C_{10})))(F^{LL}_{ef_2}+F^{LL}_{af_2}+F^{LL}_{aP})\nonumber\\
&+&(\frac{C_5}{3}+C_6-\frac{1}{2}(\frac{C_7}{3}+C_8))(F^{SP}_{ef_2}+F^{SP}_{af_2}+F^{SP}_{aP})+(C_3+2C_4-\frac{C_9}{2}+\frac{C_{10}}{2})(M^{LL}_{ef_2}\nonumber\\
&+&M^{LR}_{af_2}+M^{LL}_{eP}+M^{LL}_{aP})+(C_5-\frac{C_7}{2})(M^{LR}_{ef_2}+M^{LR}_{af_2}+M^{LR}_{eP}+M^{LR}_{aP})\nonumber\\
&+&(2C_6+\frac{C_8}{2})(M^{SP}_{ef_2}+M^{SP}_{af_2}+M^{SP}_{eP}+M^{SP}_{aP})]\big\}\;,\label{amp7}
\end{eqnarray}
 \begin{eqnarray}
{\cal A}(B^0 \to \eta_s(f_2 \to)\pi^+ \pi^-) &=& \frac{G_F} {2}
  \big\{-V_{tb}^*V_{td}[(C_3+\frac{C_4}{3}-C_5-\frac{C_6}{3}+\frac{1}{2}(C_7+\frac{C_8}{3}-C_9-\frac{C_{10}}{3}))F^{LL}_{ef_2}\nonumber\\
&+&(C_4-\frac{C_{10}}{2})M^{LL}_{ef_2}+(C_6-\frac{C_8}{2})M^{SP}_{ef_2}]\big\} \;,\label{amp8}
\end{eqnarray}
 \begin{eqnarray}
{\cal A}(B^0 \to \eta(f_2\to)\pi^+ \pi^-) &=&{\cal A}(B^0 \to f_2 \eta_q) \cos{\phi}-{\cal A}(B^0 \to f_2 \eta_s)\sin{\phi} \;,\label{amp9}\\
 {\cal A}(B^0 \to \eta^{\prime}(f_2\to)\pi^+ \pi^-) &=& {\cal A}(B^0 \to f_2 \eta_q)\sin{\phi}+{\cal A}(B^0 \to f_2 \eta_s) \cos{\phi} \;,\label{amp10}
\end{eqnarray}
 \begin{eqnarray}
{\cal A}(B_s^0 \to \eta_q(f_2 \to)\pi^+ \pi^-) &=& \frac{G_F} {2\sqrt{2}}
\big\{V_{ub}^*V_{us}[(C_1+\frac{C_2}{3})(F^{LL}_{af_2}+F^{LL}_{aP})+C_2(M^{LL}_{af_2}+M^{LL}_{aP})]\nonumber\\
&-&V_{tb}^*V_{ts}[(2C_3+\frac{2C_4}{3}-2C_5-\frac{2C_6}{3}-\frac{C_7}{2}-\frac{C_8}{6}+\frac{C_9}{2}+\frac{C_{10}}{6})(F^{LL}_{af_2}+F^{LL}_{aP})\nonumber\\
&+&(2C_4+\frac{C_{10}}{2})(M^{LL}_{af_2}+M^{LL}_{aP})+(2C_6+\frac{C_8}{2})(M^{SP}_{af_2}+M^{SP}_{aP})]\big\} \;, \label{amp11}
\end{eqnarray}
 \begin{eqnarray}
{\cal A}(B_s^0 \to \eta_s(f_2 \to)\pi^+ \pi^-) &=& \frac{G_F} {2}
\big\{V_{ub}^*V_{us}[C_2M^{LL}_{eP}]-V_{tb}^*V_{ts}[(2C_4+\frac{ C_{10}}{2})M^{LL}_{eP}+(2C_6+\frac{C_8}{2})M^{SP}_{eP}]\big\} \;,\label{amp12}\\
{\cal A}(B_s^0 \to \eta(f_2\to)\pi^+ \pi^-) &=& {\cal A}(B_s^0 \to f_2 \eta_q) \cos{\phi}-{\cal A}(B_s^0 \to f_2 \eta_s)\sin{\phi} \;,\label{amp13}\\
{\cal A}(B_s^0 \to \eta^{\prime}(f_2\to)\pi^+ \pi^-) &=& {\cal A}(B_s^0 \to f_2 \eta_q)\sin{\phi}+{\cal A}(B_s^0 \to f_2 \eta_s) \cos{\phi}  \;,
\label{amp14}
\end{eqnarray}

On the other hand, the meson $f_2(1270)$ is more like an pure $s\bar{s}$
state, the total decay amplitude for each considered decay mode can
be written  as:
\begin{eqnarray}
{\cal A}(B^+ \to K^+(f_2 \to)\pi^+ \pi^-) &=& \frac{G_F} {\sqrt{2}} \big\{V_{ub}^*V_{us}[(\frac{C_1}{3}+C_2)F^{LL}_{aP}+C_1M^{LL}_{aP}]
-V_{tb}^*V_{ts}[(C_3+C_4-\frac{1}{2}(C_9+C_{10}))M^{LL}_{eP}\nonumber\\
&+&(C_5-\frac{C_7}{2})M^{LR}_{eP}+(C_6-\frac{C_8}{2})M^{SP}_{eP}+(\frac{C_3}{3}+C_4+\frac{C_9}{3}+C_{10})F^{LL}_{aP}\nonumber\\
&+&(\frac{C_5}{3}+C_6+\frac{C_7}{3}+C_8)F^{SP}_{aP}+(C_3+C_9) M^{LL}_{aP}+(C_5+C_7)M^{LR}_{aP}]\big\} \;, \label{amp15}
\end{eqnarray}
\begin{eqnarray}
{\cal A}(B^0 \to K^0(f_2 \to)\pi^+ \pi^-) &=& \frac{G_F} {\sqrt{2}} \big\{-V_{tb}^*V_{ts}[(C_3+C_4-\frac{1}{2}(C_9+C_{10}))M^{LL}_{eP}+(C_5-\frac{C_7}{2})M^{LR}_{eP}\nonumber\\
&+&(C_6-\frac{C_8}{2})M^{SP}_{eP}+(\frac{C_3}{3}+C_4-\frac{1}{2}(\frac{C_9}{3}+C_{10}))F^{LL}_{aP}\nonumber\\
&+&(\frac{C_5}{3}+C_6-\frac{1}{2}(\frac{C_7}{3}+C_8))F^{SP}_{aP}+(C_3-\frac{C_9}{2}) M^{LL}_{aP}+(C_5-\frac{C_7}{2})M^{LR}_{aP}]\big\}\;,\label{amp16}
 \end{eqnarray}
\begin{eqnarray}
{\cal A}(B_s^0 \to \bar{K}^0(f_2 \to)\pi^+ \pi^-) &=& \frac{G_F} {\sqrt{2}}\big\{-V_{tb}^*V_{td}[(\frac{C_3}{3}+C_4-\frac{1}{2}(\frac{C_9}{3}+C_{10}))(F^{LL}_{ef_2}+F^{LL}_{af_2})\nonumber\\
&+&(\frac{C_5}{3}+C_6-\frac{1}{2}(\frac{C_7}{3}+C_8))(F^{SP}_{ef_2}+F^{SP}_{af_2})+(C_3-\frac{C_9}{2})(M^{LL}_{ef_2}+M^{LL}_{af_2})\nonumber\\
&+&(C_5-\frac{C_7}{2})(M^{LR}_{ef_2}+M^{LR}_{af_2})+(C_4-\frac{C_{10}}{2})M^{LL}_{eP}+(C_6-\frac{C_8}{2})M^{SP}_{eP}]\big\} \;,\label{amp17}
\end{eqnarray}
\begin{eqnarray}
{\cal A}(B^+ \to \pi^+(f_2 \to)\pi^+ \pi^-) &=& \frac{G_F} {\sqrt{2}}\big\{-V_{tb}^*V_{td}[(C_4-\frac{C_{10}}{2})M^{LL}_{eP}+(C_6-\frac{C_8}{2})M^{SP}_{eP}]\big\} \;,\label{amp18}
\end{eqnarray}
\begin{eqnarray}
 {\cal A}(B^0 \to \pi^0(f_2 \to)\pi^+ \pi^-) &=& -\frac{G_F} {2}\big\{-V_{tb}^*V_{td}[(C_4-\frac{C_{10}}{2})M^{LL}_{eP}+(C_6-\frac{C_8}{2})M^{SP}_{eP}]\big\} \;,\label{amp19}
\end{eqnarray}
 \begin{eqnarray}
{\cal A}(B_s^0 \to \pi^0(f_2 \to)\pi^+ \pi^-) &=& \frac{G_F}{2}
\big\{V_{ub}^*V_{us}[(C_1+\frac{C_2}{3})F^{LL}_{ef_2}+C_2M^{LL}_{ef_2}]
-V_{tb}^*V_{ts}[\frac{3}{2}(C_9+\frac{C_{10}}{3}-C_7-\frac{C_8}{3})F^{LL}_{ef_2}\nonumber\\
&+&\frac{3C_{10}}{2}M^{LL}_{ef_2}+\frac{3C_8}{2}M^{SP}_{ef_2}]\big\} \;,\label{amp20}
\end{eqnarray}
\begin{eqnarray}
{\cal A}(B^0 \to \eta_q(f_2 \to)\pi^+ \pi^-) &=& \frac{G_F} {2}\big\{-V_{tb}^*V_{td}[(C_4-\frac{C_{10}}{2})M^{LL}_{eP}+(C_6-\frac{C_8}{2})M^{SP}_{eP}]\big\}\;,\label{amp21}
\end{eqnarray}
 \begin{eqnarray}
{\cal A}(B^0 \to \eta_s(f_2 \to)\pi^+ \pi^-) &=& \frac{G_F} {\sqrt{2}}
  \big\{-V_{tb}^*V_{td}[(C_3+\frac{C_4}{3}-C_5-\frac{C_6}{3}+\frac{1}{2}(C_7+\frac{C_8}{3}-C_9-\frac{C_{10}}{3}))(F^{LL}_{af_2}+F^{LL}_{aP})\nonumber\\
&+&(C_4-\frac{C_{10}}{2})(M^{LL}_{af_2}+M^{LL}_{aP})+(C_6-\frac{C_8}{2})(M^{SP}_{af_2}+M^{SP}_{aP})]\big\} \;,\label{amp22}
\end{eqnarray}
 \begin{eqnarray}
{\cal A}(B^0 \to \eta(f_2\to)\pi^+ \pi^-) &=&{\cal A}(B^0 \to f_2 \eta_q) \cos{\phi}-{\cal A}(B^0 \to f_2 \eta_s)\sin{\phi} \;,\label{amp23}\\
 {\cal A}(B^0 \to \eta^{\prime}(f_2\to)\pi^+ \pi^-) &=& {\cal A}(B^0 \to f_2 \eta_q)\sin{\phi}+{\cal A}(B^0 \to f_2 \eta_s) \cos{\phi} \;,\label{amp24}
\end{eqnarray}
 \begin{eqnarray}
{\cal A}(B_s^0 \to \eta_q(f_2 \to)\pi^+ \pi^-) &=& \frac{G_F} {2}
\big\{V_{ub}^*V_{us}[(C_1+\frac{C_2}{3})F^{LL}_{ef_2}+C_2M^{LL}_{ef_2}]-V_{tb}^*V_{ts}[(2C_3+\frac{2C_4}{3}-2C_5-\frac{2C_6}{3}\nonumber\\
&-&\frac{C_7}{2}-\frac{C_8}{6}+\frac{C_9}{2}+\frac{C_{10}}{6})F^{LL}_{ef_2}+(2C_4+\frac{C_{10}}{2})M^{LL}_{ef_2}+(2C_6+\frac{C_8}{2})M^{SP}_{ef_2}]\big\} \;,\label{amp25}
\end{eqnarray}
 \begin{eqnarray}
{\cal A}(B_s^0 \to \eta_s(f_2 \to)\pi^+ \pi^-) &=& \frac{G_F} {\sqrt{2}}
\big\{-V_{tb}^*V_{ts}[(\frac{4}{3}(C_3+C_4-\frac{C_9}{2}-\frac{C_{10}}{2})-C_5-\frac{C_6}{3}+\frac{C_7}{2}+\frac{C_8}{6})(F^{LL}_{ef_2}+F^{LL}_{af_2}+F^{LL}_{aP})\nonumber\\
&+&(\frac{C_5}{3}+C_6-\frac{C_7}{6}-\frac{C_8}{2})(F^{SP}_{ef_2}+F^{SP}_{af_2}+F^{SP}_{aP})+(C_3+C_4-\frac{1}{2}(C_9+C_{10}))(M^{LL}_{ef_2}+M^{LL}_{af_2}\nonumber\\
&+&M^{LL}_{eP}+M^{LL}_{aP})+(C_5-\frac{C_7}{2})(M^{LR}_{ef_2}+M^{LR}_{af_2}+M^{LR}_{eP}+M^{LR}_{aP})\nonumber\\
&+&(C_6-\frac{C_8}{2})(M^{SP}_{ef_2}+M^{SP}_{af_2}+M^{SP}_{eP}+M^{SP}_{aP})]\big\} \;,\label{amp26}\\
{\cal A}(B_s^0 \to \eta(f_2\to)\pi^+ \pi^-) &=& {\cal A}(B_s^0 \to f_2 \eta_q) \cos{\phi}-{\cal A}(B_s^0 \to f_2 \eta_s)\sin{\phi} \;,\label{amp27}\\
{\cal A}(B_s^0 \to \eta^{\prime}(f_2\to)\pi^+ \pi^-) &=& {\cal A}(B_s^0 \to f_2 \eta_q)\sin{\phi}+{\cal A}(B_s^0 \to f_2 \eta_s) \cos{\phi}  \;,
\label{amp28}
\end{eqnarray}
where $G_F$ is the Fermi coupling constant.  $V_{ij}$'s are the Cabibbo-Kobayashi-Maskawa matrix elements.
The functions $ ( F^{LL}_{ef_2}, F^{LL}_{af_2}, M^{LL}_{ef_2}, M^{LL}_{af_2}, \cdots ) $
appeared in above equations are the individual decay amplitudes corresponding to different currents,
and their explicit expressions can be found in the Appendix of Ref.~\cite{plb763-29}.


\end{document}